 \documentclass[final,5p,times,twocolumn]{elsarticle}

\usepackage{amssymb}
\usepackage{bm}%
\usepackage[colorlinks=true,linkcolor=blue]{hyperref}%
\usepackage{dcolumn}%
\usepackage{graphicx}
\expandafter\ifx\csname package@font\endcsname\relax\else
 \expandafter\expandafter
 \expandafter\usepackage
 \expandafter\expandafter
 \expandafter{\csname package@font\endcsname}%
\fi
\usepackage{subcaption}
\usepackage[english]{babel}
\usepackage[numbers]{natbib}
\biboptions{numbers,sort&compress}

\begin{document}

\begin{frontmatter}



\title{Viscoelasticity and dynamical gaps: rigidity in crystallization and glass-forming liquids}%

 \author[1]{J. Quetzalc\'oatl Toledo-Mar\'in}
\author[1]{Gerardo G. Naumis  \corref{cor1} } \ead{naumis@fisica.unam.mx}

\cortext[cor1]{Corresponding author}
\address[1]{Departamento de Sistemas Complejos, Instituto de
F\'{i}sica, Universidad Nacional Aut\'{o}noma de M\'{e}xico (UNAM),
Apartado Postal 20-364, 01000 M\'{e}xico, Distrito Federal,
M\'{e}xico}






\begin{abstract}
Rigidity plays an important role on the relaxation properties of glass forming melts, yet it is usually determined from the average coordination number through the chemical composition. A discussion is presented on how viscoelasticity can be used as an alternative way to determine glass rigidity and to give clues about the relaxation processes.  It is shown that the transverse current dynamical structure factor of dense glass and crystal forming fluids contain rich information about rigidity that can be related with the presence of a dynamical-gap for transversal vibrational-modes. Then, the number of floppy modes can be related with the dynamical gap size  and with the liquid relaxation time. Furthermore, a dynamical average effective coordination number can be defined. Numerical simulations  for hard-disks in a dense fluid phase are provided.  A discussion is presented on the need to improve  glass viscoelasticity models to describe consistently non-exponential stress and strain relaxation.
\end{abstract}


\begin{keyword}
viscoelasticity \sep relaxation \sep rigidity
\PACS

\end{keyword}

\end{frontmatter}



One of the most important problems in glass formation is the understanding of structural relaxation mechanisms near glass transition \cite{Gupta2009,Zanotto2011,binder2011glassy, berthier2011theoretical, wolynes2012structural,naumisbifurcation,Gupta2016}, as well as  how  supercooled liquid relaxation wins over crystal nucleation \cite{Zanotto2018}.  Certainly a huge body of research has been focused on the subject (see \cite{pedersen2016thermodynamics,albert2016fifth, hansen2017connection,gleim2000relaxation, mezard2012glasses, trachenko2011heat, micoulaut1999glass, Mauro2} and in particular \cite{dyre2006col} and references therein), yet there is not a definitive consensus.  As is well known, experiments and simulations still have many feats to achieve \cite{ninarello2017models}.

Relaxation is related with one of the key features of glass formation:  the minimal speed required in order to make a glass, a property known as glass forming ability \cite{binder2011glassy}.  Phillips and Thorpe's rigidity theory  gives good insights on how this problem is related to network topology \cite{phillips1979topology,thorpe1983continuous}. These ideas can be extended to include non-directional potentials \cite{huerta,huerta2002evidence,flores2012mean}. Eventually, the pioneering work of Gupta and Mauro \cite{Gupta2009} led ridigidty theory to produce a new and highly accurate viscosity model, known in the literature as the Mauro-Yue-Ellison-Gupta-Allison Model (MYEGA) \cite{MYEGA}. This allows us to understand the chemical composition and temperature effects on the viscosity of glass-former melts  \cite{MYEGA}.  As a result, we are closer than ever to an age of glasses obtained by design \cite{mauro2014glass, siqueira2017bioglass} . This goes together with the advances made by Micoulaut and Bauchy who had extensively studied how to define rigidity for realistic potentials (see for instance \cite{bauchy2011atomic}). Previous efforts were made in simple models by Huerta et. al.\cite{huerta,huerta2002evidence,Huerta2003,Huerta2004}. Stochastic models also provided a different pathway to include chemical composition effects \cite{naumis1998stochastic,kerner2000stochastic}. From an experimental point of view, Boolchand and coworkers have extensively studied the optical, mechanical and thermodynamical properties in terms of rigidity \cite{selvanathan2000stiffness, gunasekera2013superstrong}. Theoretical models allows an understanding of some general properties of thermodynamics in terms of rigidity \cite{naumis2005energy,yan2018entropy} and there are suggestions of a connection with the boson peak \cite{Flores2009,Flores2011,Flores2010}.

As a matter of fact,  any symmetry-breaking thermodynamic phase transition involves the development of some kind of generalized rigidity by the system  \cite{LubenskyBook}. This allows the given system to preserve the phase order against thermal fluctuations \cite{LubenskyBook}. 
 In spite of this fundamental character, it is surprising to find that  in general such observation is not emphasized when phase transitions are studied. A fluid is different from a solid precisely due to its rigidity,  and thus a simple first-order fluid-solid phase transition  must also contain a rigidity transition as its main signature. Moreover, the lack of rigidity is the defining property of a Newtonian fluid, i.e., the absence of elastic behavior against shear stress. This leads to the absence of transversal waves in a fluid. 
 
 The main aim of this work is to emphasize the need to decode how the rigidity of glass forming melts  depends upon the time and spatial scales in which the system is probed or perturbed. Moreover, to accomplish this feat we need to understand rigidity transitions not only in glasses, but for crystallization and in cluster nucleation at the kinetic spinodal temperature \cite{ZanottoMauro}. 
 
 A lot of knowledge on these aspects could be obtained by looking at the similarities and differences between rigidity in organic and inorganic glasses, all of them above $T_g$ and close to the mechanical vitrification point \cite{Bartenev}. For organic glasses there is a vast literature concerning flexible and rigid polymer models in which relaxation has been tested \cite{Bartenev,Wiener1,Wiener2}. On the experimental side, modulated differential scanning calorimetry (MDSC), dielectric relaxation and rehology measurements have been very succesful for the understanding of relaxation processes \cite{macosko1994rheology}. For inorganic glasses, a series of different experiments such as MDSC and Raman scattering among others, as well as computational simulations reveal interesting aspects of the rigidity transition \cite{selvanathan2000stiffness, holbrook2014topology, yildirim2016universal, boolchand2018topological}. However, results on chalcogenide glasses frequency-dependent rheology are recent \cite{gueguen2015,SulfurPRL,Zhou2018,Sen}. 
 
 One may wonder what is the fundamental difference between rigidity in organic and inorganic glasses above $T_g$. Many years ago the answer to this question was not clear. Above $T_g$ and due to their polymeric nature, organic glasses display transitions from the folded to the stretched chain forms and thus present viscoelasticity \cite{macosko1994rheology}. On the other hand, inorganic glass melts were thought to be purely Newtonian fluids  \cite{Bartenev}. Yet, inorganic glasses, as polymeric systems, were expected to display viscoelasticity.  This apparent paradox was solved by the observation made by G. M. Bartenev, who started by adscribing the prominent differences between the $T_g$ of inorganic and organic glasses to the much higher rotation flexibility of the C-C bonds \cite{Bartenev}. Therefore, it was concluded that viscoelasticity was also possible for inorganic glasses, although happens to be smaller than in its organic counterparts due to their somewhat limited angular bond excursions. This sole fact explained why the viscoelastic response in inorganic glasses remained for a long time unnoticed  \cite{Bartenev}.

 For organic glasses, the key to understand the relationship between relaxation and rigidity is given by measuring the viscoelasticity using rheological experiments \cite{macosko1994rheology}. In viscoelasticity, the relationship between the stress $\sigma (\omega)$ and strain $\epsilon (\omega)$ is measured as a function of the frequency $\omega$. For $\epsilon(\omega)=\cos \omega t$, we have $\sigma(\omega)=G'\cos \omega t+G''\sin \omega t $,  and thus a complex modulus $G(\omega)$ is obtained \cite{SulfurPRL}. The real part of $G(\omega)$, denoted by  $G'(\omega)$, is the storage modulus while the imaginary part $G''(\omega)$ gives the loss modulus. The phase lag between strain and stress is given by $\tan \theta (\omega)=G''(\omega)/G'(\omega)$, while a frequency-dependent viscosity is obtained from $\eta(\omega)=\sqrt{G''(\omega)^2+G'(\omega)^2}/\omega$. 
 
 Above the glass transition and for low frequencies\cite{gueguen2015,SulfurPRL,Zhou2018,Sen}, the viscosity $\eta(\omega)$ is strongly frequency-dependent
 and $G'(\omega)<<G''(\omega)$. At these frequencies, the system behaves as a viscous fluid as $\eta(\omega) \approx G''(\omega)/\omega$.
  In the limit  of high frequencies, denoted by $\omega=\infty$, we have that $G(\infty) \approx G'(\infty) > G''(\infty)$ and mainly a purely elastic response is obtained. The lesson taken from these rheology experiments is that at high-frequencies, the system is rigid-like, while at low frequencies is non-rigid. Thus, rigidity in melts involves the time-scale in which the system is probed. Moreover, this aspect means that there must also be a transition concerning the propagation of transversal waves. As the dispersion relationship of waves involves $\omega$ as a function of the wavevector $\boldsymbol{k}$, is clear that rigidity involves time and space density-density fluctuations.  
 A striking demonstration of this phenomena is the report of  transversal-wave branchs in the dynamical structure factor \cite{boon1991molecular,trachenko2015collective, huerta2015collective}. The transversal part of the dynamical structure factor is defined as \cite{boon1991molecular},
 \begin{equation}
 S(k,\omega)= \int_0^{\infty} dt e^{-\imath \omega t} C(k,t) \; .
 \end{equation}   
 where $C(k,t)$ is the transversal current density correlation function,
\begin{equation}
C(k,t)= \langle  J_T^*(k,t) J_T(k,0) \rangle \; , \label{eq:C(k,t)}
\end{equation} 
and the brackets $\langle ... \rangle $ represent an ensemble average. The function $J_T(k,t)$ is the transversal density current averaged over the different directions of $\boldsymbol{k}$ given the wavenumber $k=\vert \boldsymbol{k} \vert$,
\begin{equation}
J_T(k,t)=\frac{1}{\sqrt{2 N}k}\sum_{i=1}^{N}\boldsymbol{k} \times \boldsymbol{v}_i(t) \exp\left(\imath \boldsymbol{k} \cdot \boldsymbol{r}_i(t) \right) \; .
\end{equation}

Here, $\boldsymbol{v}_i(t)$ and $\boldsymbol{r}_i(t)$ are the velocity and position of the $i^{th}$ particle of a given system at time $t$. The $1/\sqrt{2}$ factor takes into account the two transverse currents in three-dimensional systems, and is replaced by one in two dimensions.

\begin{figure}[hbtp]
\centering
\includegraphics[width=3.4in]{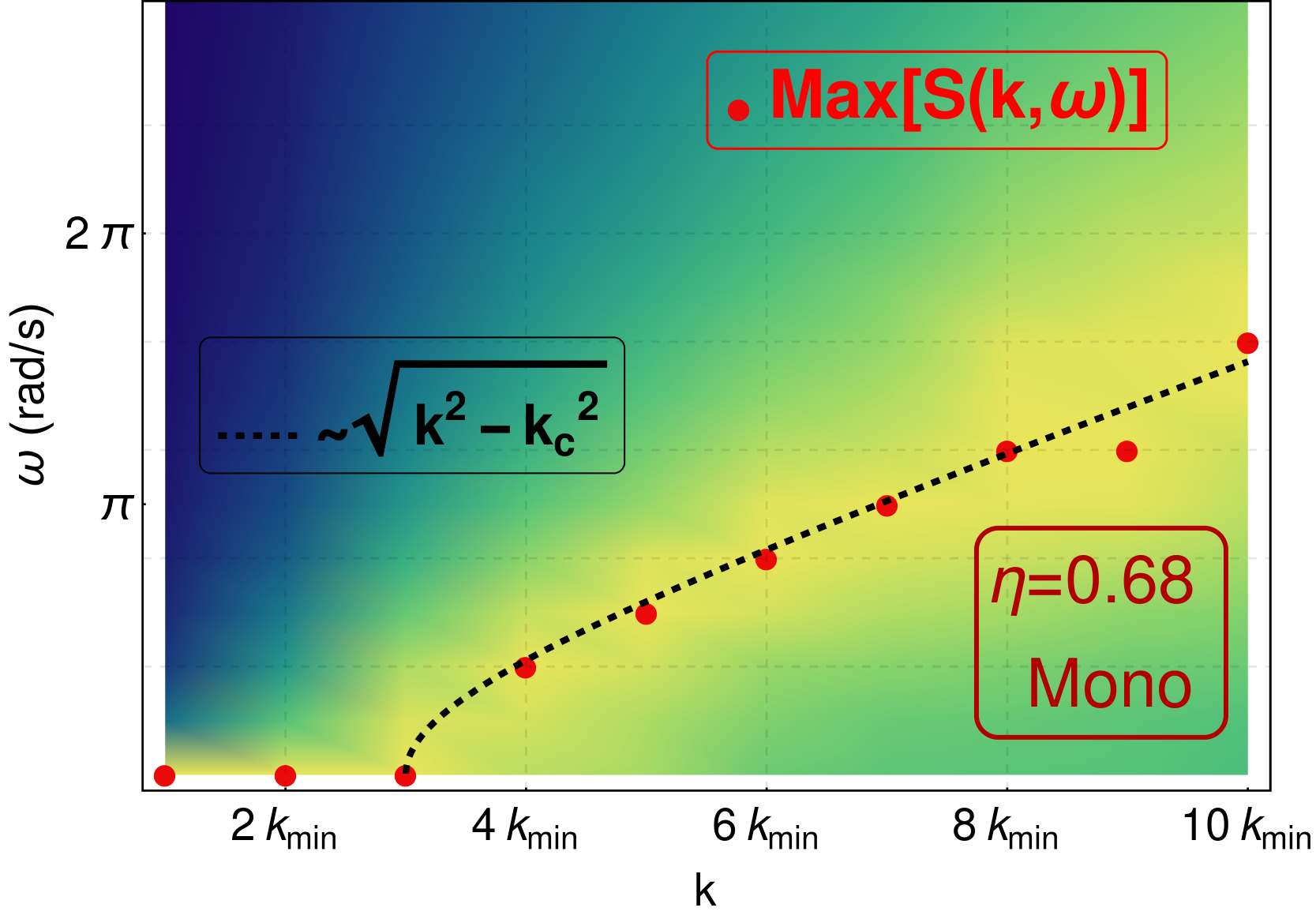}
\includegraphics[width=3.5in]{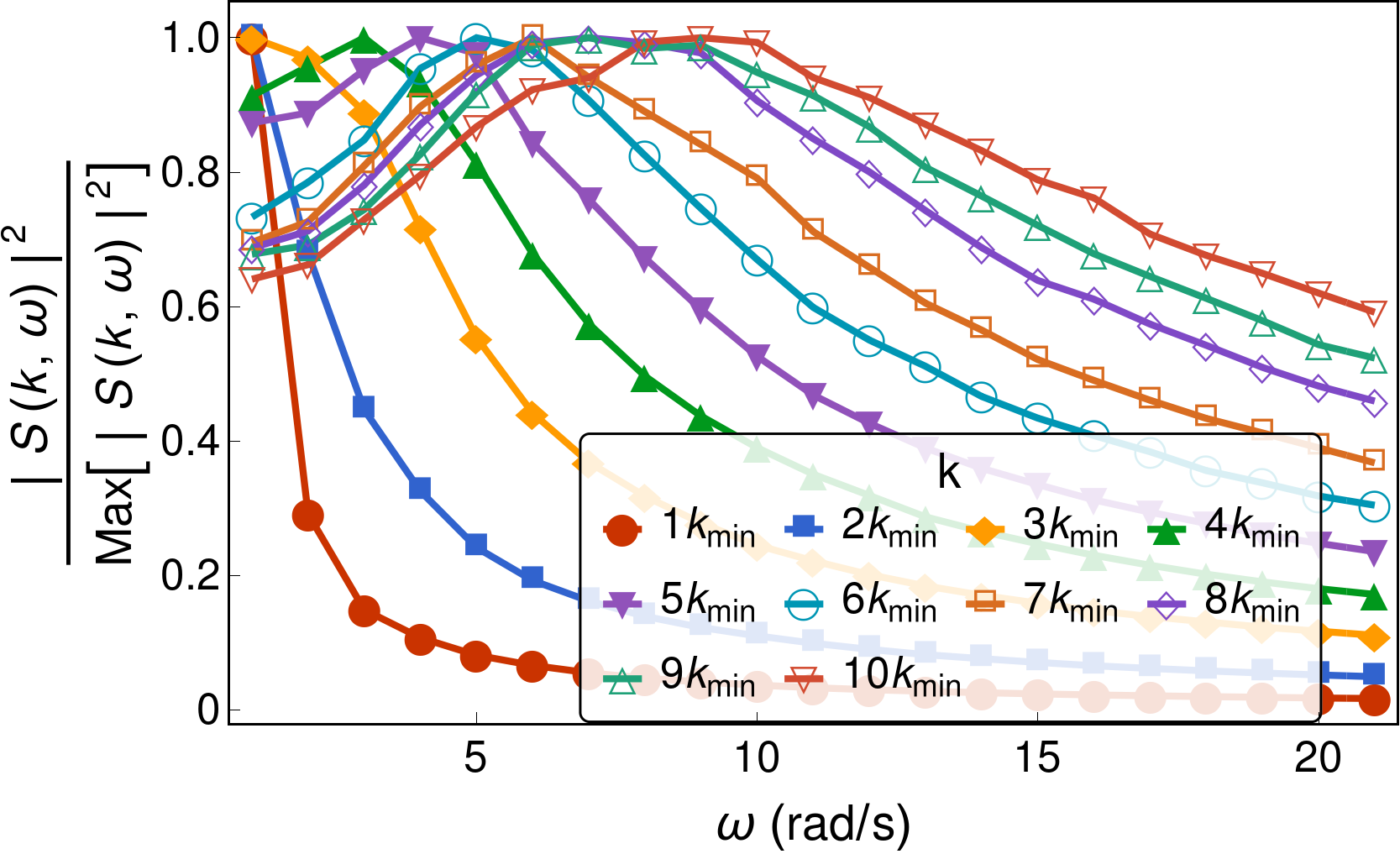}
\caption{The transversal part of the dynamical structure factor $S(k,\omega)$ in a system of $2500$ monodisperse hard disks with periodic boundary conditions and packing fraction $\phi=0.68$, which is in the fluid phase but close to the threshold where the system freezes. \textbf{Upper Panel}: Contour plot of the normalized transversal part of the dynamical structure factor as a function of $\omega$ and the wavenumber $k$, given in terms of $k_{min}=4 \sqrt{\eta \pi/N}$. The red points correspond to the maximal dynamical structure factor values, and the lines through them are visual guides. The dashed curve $\sqrt{k^2-k_c^2}$ is presented for comparison purposes.   \textbf{Lower Panel}: The transversal part of the dynamic Structure Factor vs $\omega$ for different wavenumbers $k$  (see legend). For wavenumbers $k$ equal to $ k_{min}$, $2 k_{min}$ and $3 k_{min}$, the transversal part of the dynamical structure factor has a peak at $\omega=0$. For wavenumber $k \geq 4 k_{min}$, the transversal part of the dynamical structure factor has peaks at $\omega\neq 0$. The dynamical $k$-gap satisfies the inequality $3k_{min}<k_c<4k_{min}$, i.e., for wave numbers. The lines connecting the plotmarkers are meant as visual guides.} \label{fig:Sqw}
\end{figure}

 As an example, in Figure \ref{fig:Sqw} we present the transversal part of the dynamical structure factor $S(k,\omega)$ for the simplest imaginable system: hard-disks. This result was obtained from a molecular dynamical simulation of $2500$ hard-disks. Once the simulation was thermalized, we ran the simulation 2000 times for different velocities and positions. The transversal current density correlation function, Eq. (\ref{eq:C(k,t)}) was averaged over these 2000 simulation samples in order to reduce the noise. We used the event driven molecular dynamics simulation called \textit{DynamO} \cite{bannerman2011dynamo}.  It is important to remark that the presented results in Figure \ref{fig:Sqw} were obtained for a packing fraction $\phi=0.68$, where the system is in a very dense fluid phase, close to the freezing point which is known to be at $\phi_C\approx 0.72$. \cite{speedy1999glass, russo2017disappearance}

In the upper panel of Fig. \ref{fig:Sqw}, we show the resulting contour plot of $S(k,\omega)$. We can see that for small wavenumbers, shear waves do not propagate as expected for the fluid phase. However, Fig. \ref{fig:Sqw} reveals
a threshold $k_c$. Whenever $k>k_c$, shear waves indeed propagate in the system. In the lower panel of Fig. \ref{fig:Sqw}, we show the transversal part of the dynamical structure factor \textit{vs} $\omega$ for different wavenumbers $k$ given in terms of the lowest wavenumber  $k_{min}=4 \sqrt{\eta \pi/N}$. Notice how as $k$ increases, the  peaks in $S(k,\omega)$ shift to larger values of $\omega$. Furthermore, there is a gap between the peaks for $k\leq3k_{min}$ and $k=4k_{min}$.  From the upper panel in Fig. \ref{fig:Sqw} we can see that $\omega(k) \approx \sqrt{k^2-k_c^2} $ in agreement with a recent theoretical solid-state approach to liquids \cite{trachenko2015collective, baggioli2018maxwell, baggioli2018solidity}.

Fig. \ref{fig:Sqw} shows another viewpoint to look at viscoelasticity, but here the change from a fluid-like to a solid-like behavior is revealed by the presence of a dynamical gap \cite{boon1991molecular,trachenko2015collective, yang2017emergence}. Transversal wave propagation is only possible for modes with $k>k_c$. For $k<k_c$, in Fig. \ref{fig:Sqw} we observe that $S(\boldsymbol{k},\omega) \approx \delta(\omega) $, where $\delta(\omega)$ is the Dirac delta function. As  for $k<k_c$ we have $\omega=0$, we can consider these states in terms of rigidity  as floppy, i.e., the system is flexible.

In general we can estimate a relationship between $k_c$ and the number of floppy modes as follows. Since the fluid is isotropic, the number floppy modes in three dimensions is,
\begin{equation}
N_f(k_c) \approx  2 \int_{0}^{k_c} 4 \pi k^{2} dk=\frac{8\pi}{3}  k_c^3
 \end{equation}
The fraction of floppy modes ($f$) with respect to the total number of modes is then,
\begin{equation}\label{f}
f \approx \frac{2}{3}\left(\frac{k_c}{k_D}\right)^3
 \end{equation} 
 The normalization factor $k_D \; (\gg k_c)$ is the Debye wavevector \cite{yang2017emergence}. We thus arrive to the conclusion that floppy modes are related with a dynamical gap. Moreover, as 
 $k_c=1/c\tau(T)$, where $c$ is the transverse sound speed and $\tau(T)$ is the  average time at temperature $T$ it takes for a molecule to diffuse a distance equal to the inter-atomic separation \cite{yang2017emergence}, we can further relate floppy modes with this characteristic time, 
 \begin{equation}
 f \approx \frac{2}{3\omega_D^3}\left(\frac{1}{\tau^3(T)}\right)
 \end{equation} 
 where $ \omega_D=ck_D$. Although floppy modes in principle reduce the internal energy \cite{naumis2006variation,yang2017emergence}, this will not happen in all cases, as entropy has two sources, vibrational and configurational \cite{naumis2005energy}.
 
As a matter of fact, floppy regions favor the maximization of vibrational entropy \cite{naumis2005energy} and thus under certain conditions domains of floppy regions appear \cite{yan2018entropy}. This in turn has huge consequences for relaxation \cite{glasstone1941theory, hanggi, toledo2017short, toledo2018escape} and it becomes  difficult to characterize rigidity using a mean-field approach above glass transition. Nevertheless, following the spirit of a mean field , we can define a mean coordination number $<r>$ of an effective topological lattice \cite{thorpe1983continuous}. The fraction of floppy modes is $f=(3N-c)/3N$, where $c$ is the number of constrains. When angular and radial forces are present, this results in $f=2-5<r>/6$, while $f=1-<r>/6$ for radial forces. 
 By using Eq. (\ref{f}) we arrive to a possible and alternative definition for a ``dynamical'' mean coordination number in the melt when angular forces are present,
 \begin{equation}
  <r>=\frac{12}{5}\left(1 -\frac{1}{3}\left(\frac{k_c}{k_D}\right)^3\right)
 \end{equation}
 and for pure radial forces,
 \begin{equation}
  <r>=6\left(1 -\frac{2}{3}\left(\frac{k_c}{k_D}\right)^3\right)
 \end{equation}

 We remark that here $k_c=0$ implies  $<r>=2.4$ whenever angular forces are present. In a similar way, $k_c=0$ implies  $<r>=6$ for pure radial forces.  These are the magical coordinations for  rigidity transitions \cite{phillips1979topology} and thus contain and highlight what we expect for a transition from a liquid to a solid. When there is a hierarchy of forces, these coordination numbers are not intended to necessarily caracterize glasses below $T_g$  as the solidified network can be already classified as floppy, isostatic or rigid. This task requires a more involved treatment,  yet the present ideas suggest a path to be followed.

Let us discuss these dynamical results in the context of the usual invoked arguments relating relaxation time ($\tau$) and Newtonian viscosity ($\eta_0$) above $T_g$. This characteristic time at glass transition is estimated  by setting $\eta_0 \approx 10^{12} \ Pa \ s$ in the Maxwell relationship $\tau \approx \eta_0/G'(\infty)$.  This comes from the simplest model of viscoelasticity: a spring with a dashpot connected in series. However, the Maxwell model  automatically implies  exponential stress relaxation \cite{gutzow1995vitreous}. Glasses and glass-forming melts are known to have non-exponential relaxation \cite{gutzow1995vitreous}, as for example,  streteched exponential relaxation $\sigma(t)=\sigma_0\exp[-t/\tau]^{\beta}$ where $\beta$ depends upon the range of the interaction \cite{naumisbifurcation}, as happens for relaxation in other topologically connected lattices \cite{naumistails,NaumisPhillips2012}.  To be consistent, is paramount to search beyond the Maxwell picture. The task can be performed by using an extensive collection of models \cite{Bartenev,gutzow1995vitreous}. Several paths are envisioned which include the use of fractional derivatives  and generalized Maxwell-Voigt-Zener models with many spring dash-pots circuits to accurately reproduce all frequency  decades \cite{sudduth2005}.  This is in agreement with the use of Prony series to represent many relaxations times in order to obtain an accurate SER \cite{MauroProny}. However, even for organic glasses is difficult to obtain models able to reproduce all kind of possible protocols for elasticity measurements  \cite{sudduth2005}. Moreover, for inorganic glasses the relationship between rigidity and elasticity protocols is still a work in progress. For chalcogenide glasses,  recent works add to viscoelasticity a plastic response \cite{Zhou2018}  or a delayed elasticity \cite{gueguen2015} to account for the results on specific protocols. Any advance in this area is essential, as elastic stresses are related to thermodynamic driving forces for crystallization \cite{AbyzovZanotto}. In this regard, Grassia \textit{et al.} have made significant progress  \cite{grassia2006constitutive, grassia2009interplay, grassia2009relative, grassia2011isobaric, grassia2011modeling2}. By linking viscoelasticity and the phenomenological KAHR model for structural relaxation, developed by Kovacs, Aklonis, Hutchinson, and Ramos \cite{ramos1984isobaric, mckenna2012physical}, they were able to characterize amorphous polymers and, in particular, predict the isobaric and isothermal glass transition for polystyrene \cite{grassia2011modeling}. 

Finally, we conclude by observing that for the system presented in Fig. \ref{fig:Sqw}, the dynamical gap goes to zero ($k_c \rightarrow 0$) as the hexatic to solid second order phase transition is approached. It remains to determine how the transition to rigidity occurs in glass forming melts, for example, by considering polydisperse disks. Also, we need to perform simulations on realistic Hamiltonians with angular dependent potentials. For organic glasses, it is known that such contributions   increase relaxation times by steric shielding \cite{Wiener2}. 
Cuts of the polymer chains and therefore, chain length,  is an important parameter for relaxation in organic glasses  \cite{Bartenev,Wiener1}, yet is a factor that still needs to be addressed in time-dependent constraint theory for inorganic glasses.

\section*{Acknowledgments}

This work was partially supported by DGAPA-UNAM project IN102717. J.Q.T.M. acknowledges a doctoral fellowship from CONACyT.

\newpage

\bibliography{GlassViscoelasticityBIB}

\begin{thebibliography}{10}
\expandafter\ifx\csname url\endcsname\relax
  \def\url#1{\texttt{#1}}\fi
\expandafter\ifx\csname urlprefix\endcsname\relax\def\urlprefix{URL }\fi
\expandafter\ifx\csname href\endcsname\relax
  \def\href#1#2{#2} \def\path#1{#1}\fi

\bibitem{Gupta2009}
P.~K. Gupta, J.~C. Mauro, \href{https://doi.org/10.1063/1.3077168}{Composition
  dependence of glass transition temperature and fragility. i. a topological
  model incorporating temperature-dependent constraints}, The Journal of
  Chemical Physics 130~(9) (2009) 094503.
\newblock \href {http://arxiv.org/abs/https://doi.org/10.1063/1.3077168}
  {\path{arXiv:https://doi.org/10.1063/1.3077168}}, \href
  {http://dx.doi.org/10.1063/1.3077168} {\path{doi:10.1063/1.3077168}}.
\newline\urlprefix\url{https://doi.org/10.1063/1.3077168}

\bibitem{Zanotto2011}
M.~L.~F. Nascimento, V.~M. Fokin, E.~D. Zanotto, A.~S. Abyzov,
  \href{https://doi.org/10.1063/1.3656696}{Dynamic processes in a silicate
  liquid from above melting to below the glass transition}, The Journal of
  Chemical Physics 135~(19) (2011) 194703.
\newblock \href {http://arxiv.org/abs/https://doi.org/10.1063/1.3656696}
  {\path{arXiv:https://doi.org/10.1063/1.3656696}}, \href
  {http://dx.doi.org/10.1063/1.3656696} {\path{doi:10.1063/1.3656696}}.
\newline\urlprefix\url{https://doi.org/10.1063/1.3656696}

\bibitem{binder2011glassy}
K.~Binder, W.~Kob, Glassy materials and disordered solids: An introduction to
  their statistical mechanics, World Scientific, 2011.

\bibitem{berthier2011theoretical}
L.~Berthier, G.~Biroli, Theoretical perspective on the glass transition and
  amorphous materials, Reviews of Modern Physics 83~(2) (2011) 587.

\bibitem{wolynes2012structural}
P.~G. Wolynes, V.~Lubchenko, Structural Glasses and Supercooled Liquids:
  Theory, Experiment, and Applications, John Wiley \& Sons, 2012.

\bibitem{naumisbifurcation}
G.~Naumis, J.~Phillips, Bifurcation of stretched exponential relaxation in
  microscopically homogeneous glasses, Journal of Non-Crystalline Solids
  358~(5) (2012) 893--897.

\bibitem{Gupta2016}
P.~K. Gupta, D.~R. Cassar, E.~D. Zanotto,
  \href{https://doi.org/10.1063/1.4964674}{Role of dynamic heterogeneities in
  crystal nucleation kinetics in an oxide supercooled liquid}, The Journal of
  Chemical Physics 145~(21) (2016) 211920.
\newblock \href {http://arxiv.org/abs/https://doi.org/10.1063/1.4964674}
  {\path{arXiv:https://doi.org/10.1063/1.4964674}}, \href
  {http://dx.doi.org/10.1063/1.4964674} {\path{doi:10.1063/1.4964674}}.
\newline\urlprefix\url{https://doi.org/10.1063/1.4964674}

\bibitem{Zanotto2018}
E.~D. Zanotto, D.~R. Cassar, \href{https://doi.org/10.1063/1.5034091}{The race
  within supercooled liquids—relaxation versus crystallization}, The Journal
  of Chemical Physics 149~(2) (2018) 024503.
\newblock \href {http://arxiv.org/abs/https://doi.org/10.1063/1.5034091}
  {\path{arXiv:https://doi.org/10.1063/1.5034091}}, \href
  {http://dx.doi.org/10.1063/1.5034091} {\path{doi:10.1063/1.5034091}}.
\newline\urlprefix\url{https://doi.org/10.1063/1.5034091}

\bibitem{pedersen2016thermodynamics}
U.~R. Pedersen, L.~Costigliola, N.~P. Bailey, T.~B. Schr{\o}der, J.~C. Dyre,
  Thermodynamics of freezing and melting, Nature communications 7 (2016) 12386.

\bibitem{albert2016fifth}
S.~Albert, T.~Bauer, M.~Michl, G.~Biroli, J.-P. Bouchaud, A.~Loidl,
  P.~Lunkenheimer, R.~Tourbot, C.~Wiertel-Gasquet, F.~Ladieu, Fifth-order
  susceptibility unveils growth of thermodynamic amorphous order in
  glass-formers, Science 352~(6291) (2016) 1308--1311.

\bibitem{hansen2017connection}
H.~W. Hansen, B.~Frick, T.~Hecksher, J.~C. Dyre, K.~Niss, Connection between
  fragility, mean-squared displacement, and shear modulus in two van der waals
  bonded glass-forming liquids, Physical Review B 95~(10) (2017) 104202.

\bibitem{gleim2000relaxation}
T.~Gleim, W.~Kob, The-relaxation dynamics of a simple liquid, The European
  Physical Journal B-Condensed Matter and Complex Systems 13~(1) (2000) 83--86.

\bibitem{mezard2012glasses}
M.~Mezard, G.~Parisi, Glasses and replicas, Structural Glasses and Supercooled
  Liquids: Theory, Experiment, and Applications (2012) 151--191.

\bibitem{trachenko2011heat}
K.~Trachenko, V.~Brazhkin, Heat capacity at the glass transition, Physical
  Review B 83~(1) (2011) 014201.

\bibitem{micoulaut1999glass}
M.~Micoulaut, G.~Naumis, Glass transition temperature variation, cross-linking
  and structure in network glasses: a stochastic approach, EPL (Europhysics
  Letters) 47~(5) (1999) 568.

\bibitem{Mauro2}
J.~C. Mauro, D.~C. Allan, M.~Potuzak, Nonequilibrium viscosity of glass,
  Physical Review B 80~(9) (2009) 094204.

\bibitem{dyre2006col}
J.~C. Dyre, Colloquium: The glass transition and elastic models of
  glass-forming liquids, Reviews of modern physics 78~(3) (2006) 953.

\bibitem{ninarello2017models}
A.~Ninarello, L.~Berthier, D.~Coslovich, Models and algorithms for the next
  generation of glass transition studies, Physical Review X 7~(2) (2017)
  021039.

\bibitem{phillips1979topology}
J.~C. Phillips, Topology of covalent non-crystalline solids i: Short-range
  order in chalcogenide alloys, Journal of Non-Crystalline Solids 34~(2) (1979)
  153--181.

\bibitem{thorpe1983continuous}
M.~Thorpe, Continuous deformations in random networks, Journal of
  Non-Crystalline Solids 57~(3) (1983) 355--370.

\bibitem{huerta}
A.~Huerta, G.~Naumis, Relationship between glass transition and rigidity in a
  binary associative fluid, Physics Letters A 299~(5) (2002) 660--665.

\bibitem{huerta2002evidence}
A.~Huerta, G.~G. Naumis, Evidence of a glass transition induced by rigidity
  self-organization in a network-forming fluid, Physical Review B 66~(18)
  (2002) 184204.

\bibitem{flores2012mean}
H.~M. Flores-Ruiz, G.~G. Naumis, Mean-square-displacement distribution in
  crystals and glasses: An analysis of the intrabasin dynamics, Physical Review
  E 85~(4) (2012) 041503.

\bibitem{MYEGA}
J.~C. Mauro, Y.~Yue, A.~J. Ellison, P.~K. Gupta, D.~C. Allan,
  \href{https://www.pnas.org/content/106/47/19780}{Viscosity of glass-forming
  liquids}, Proceedings of the National Academy of Sciences 106~(47) (2009)
  19780–19784.
\newblock \href
  {http://arxiv.org/abs/https://www.pnas.org/content/106/47/19780.full.pdf}
  {\path{arXiv:https://www.pnas.org/content/106/47/19780.full.pdf}}, \href
  {http://dx.doi.org/10.1073/pnas.0911705106}
  {\path{doi:10.1073/pnas.0911705106}}.
\newline\urlprefix\url{https://www.pnas.org/content/106/47/19780}

\bibitem{mauro2014glass}
J.~C. Mauro, C.~S. Philip, D.~J. Vaughn, M.~S. Pambianchi, Glass science in the
  united states: current status and future directions, International Journal of
  Applied Glass Science 5~(1) (2014) 2--15.

\bibitem{siqueira2017bioglass}
R.~L. Siqueira, L.~C. Costa, M.~A. Schiavon, D.~T. de~Castro, A.~C. dos Reis,
  O.~Peitl, E.~D. Zanotto, Bioglass{\textregistered} and resulting crystalline
  materials synthesized via an acetic acid-assisted sol--gel route, Journal of
  Sol-Gel Science and Technology 83~(1) (2017) 165--173.

\bibitem{bauchy2011atomic}
M.~Bauchy, M.~Micoulaut, Atomic scale foundation of temperature-dependent
  bonding constraints in network glasses and liquids, Journal of
  Non-Crystalline Solids 357~(14) (2011) 2530--2537.

\bibitem{Huerta2003}
A.~Huerta, G.~G. Naumis,
  \href{https://link.aps.org/doi/10.1103/PhysRevLett.90.145701}{Role of
  rigidity in the fluid-solid transition}, Phys. Rev. Lett. 90 (2003) 145701.
\newblock \href {http://dx.doi.org/10.1103/PhysRevLett.90.145701}
  {\path{doi:10.1103/PhysRevLett.90.145701}}.
\newline\urlprefix\url{https://link.aps.org/doi/10.1103/PhysRevLett.90.145701}

\bibitem{Huerta2004}
A.~Huerta, G.~G. Naumis, D.~T. Wasan, D.~Henderson, A.~Trokhymchuk,
  \href{https://doi.org/10.1063/1.1632893}{Attraction-driven disorder in a
  hard-core colloidal monolayer}, The Journal of Chemical Physics 120~(3)
  (2004) 1506--1510.
\newblock \href {http://arxiv.org/abs/https://doi.org/10.1063/1.1632893}
  {\path{arXiv:https://doi.org/10.1063/1.1632893}}, \href
  {http://dx.doi.org/10.1063/1.1632893} {\path{doi:10.1063/1.1632893}}.
\newline\urlprefix\url{https://doi.org/10.1063/1.1632893}

\bibitem{naumis1998stochastic}
G.~G. Naumis, R.~Kerner, Stochastic matrix description of glass transition in
  ternary chalcogenide systems, Journal of non-crystalline solids 231~(1)
  (1998) 111--119.

\bibitem{kerner2000stochastic}
R.~Kerner, G.~G. Naumis, Stochastic matrix description of the glass transition,
  Journal of Physics: Condensed Matter 12~(8) (2000) 1641.

\bibitem{selvanathan2000stiffness}
D.~Selvanathan, W.~Bresser, P.~Boolchand, Stiffness transitions in si x se 1- x
  glasses from raman scattering and temperature-modulated differential scanning
  calorimetry, Physical Review B 61~(22) (2000) 15061.

\bibitem{gunasekera2013superstrong}
K.~Gunasekera, S.~Bhosle, P.~Boolchand, M.~Micoulaut, Superstrong nature of
  covalently bonded glass-forming liquids at select compositions, The Journal
  of chemical physics 139~(16) (2013) 164511.

\bibitem{naumis2005energy}
G.~G. Naumis, Energy landscape and rigidity, Physical Review E 71~(2) (2005)
  026114.

\bibitem{yan2018entropy}
L.~Yan, Entropy favors heterogeneous structures of networks near the rigidity
  threshold, Nature communications 9~(1) (2018) 1359.

\bibitem{Flores2009}
H.~M. Flores-Ruiz, G.~G. Naumis,
  \href{https://doi.org/10.1063/1.3246805}{Excess of low frequency vibrational
  modes and glass transition: A molecular dynamics study for soft spheres at
  constant pressure}, The Journal of Chemical Physics 131~(15) (2009) 154501.
\newblock \href {http://arxiv.org/abs/https://doi.org/10.1063/1.3246805}
  {\path{arXiv:https://doi.org/10.1063/1.3246805}}, \href
  {http://dx.doi.org/10.1063/1.3246805} {\path{doi:10.1063/1.3246805}}.
\newline\urlprefix\url{https://doi.org/10.1063/1.3246805}

\bibitem{Flores2011}
H.~M. Flores-Ruiz, G.~G. Naumis,
  \href{https://link.aps.org/doi/10.1103/PhysRevB.83.184204}{Boson peak as a
  consequence of rigidity: A perturbation theory approach}, Phys. Rev. B 83
  (2011) 184204.
\newblock \href {http://dx.doi.org/10.1103/PhysRevB.83.184204}
  {\path{doi:10.1103/PhysRevB.83.184204}}.
\newline\urlprefix\url{https://link.aps.org/doi/10.1103/PhysRevB.83.184204}

\bibitem{Flores2010}
H.~M. Flores-Ruiz, G.~G. Naumis, J.~C. Phillips,
  \href{https://link.aps.org/doi/10.1103/PhysRevB.82.214201}{Heating through
  the glass transition: A rigidity approach to the boson peak}, Phys. Rev. B 82
  (2010) 214201.
\newblock \href {http://dx.doi.org/10.1103/PhysRevB.82.214201}
  {\path{doi:10.1103/PhysRevB.82.214201}}.
\newline\urlprefix\url{https://link.aps.org/doi/10.1103/PhysRevB.82.214201}

\bibitem{LubenskyBook}
P.~M. Chaikin, T.~C. Lubensky, Principles of Condensed Matter Physics,
  Cambridge University Press, Cambridge, 1995.

\bibitem{ZanottoMauro}
E.~D. Zanotto, J.~Mauro, The glassy state of matter: Its definition and
  ultimate fate, Journal of Non-Crystalline Solids 471 (2017) 490--495.
\newblock \href {http://dx.doi.org/10.1016/j.jnoncrysol.2017.05.019}
  {\path{doi:10.1016/j.jnoncrysol.2017.05.019}}.

\bibitem{Bartenev}
G.~M. Bartenev, \href{https://nla.gov.au/nla.cat-vn1842910}{The structure and
  mechanical properties of inorganic glasses. By G. M. Bartenev. Translated by
  [P. Jaray and] F. F. Jaray}, Wolters-Noordhoff Groningen, 1970.
\newline\urlprefix\url{https://nla.gov.au/nla.cat-vn1842910}

\bibitem{Wiener1}
R.~C. Picu, J.~H. Weiner, \href{https://doi.org/10.1063/1.475907}{Stress
  relaxation in a diatomic liquid}, The Journal of Chemical Physics 108~(12)
  (1998) 4984--4991.
\newblock \href {http://arxiv.org/abs/https://doi.org/10.1063/1.475907}
  {\path{arXiv:https://doi.org/10.1063/1.475907}}, \href
  {http://dx.doi.org/10.1063/1.475907} {\path{doi:10.1063/1.475907}}.
\newline\urlprefix\url{https://doi.org/10.1063/1.475907}

\bibitem{Wiener2}
R.~C. Picu, G.~Loriot, J.~H. Weiner,
  \href{https://doi.org/10.1063/1.478351}{Toward a unified view of stress in
  small-molecular and in macromolecular liquids}, The Journal of Chemical
  Physics 110~(9) (1999) 4678--4686.
\newblock \href {http://arxiv.org/abs/https://doi.org/10.1063/1.478351}
  {\path{arXiv:https://doi.org/10.1063/1.478351}}, \href
  {http://dx.doi.org/10.1063/1.478351} {\path{doi:10.1063/1.478351}}.
\newline\urlprefix\url{https://doi.org/10.1063/1.478351}

\bibitem{macosko1994rheology}
C.~W. Macosko, R.~G. Larson, Rheology: principles, measurements, and
  applications, Vch New York, 1994.

\bibitem{holbrook2014topology}
C.~Holbrook, S.~Chakraborty, S.~Ravindren, P.~Boolchand, J.~T. Goldstein,
  C.~Stutz, Topology and glass structure evolution in (bao) x ((b2o3) 32 (sio2)
  68) 100- x ternary—evidence of rigid, intermediate, and flexible phases,
  The Journal of Chemical Physics 140~(14) (2014) 144506.

\bibitem{yildirim2016universal}
C.~Yildirim, M.~Micoulaut, P.~Boolchand, I.~Kantor, O.~Mathon, J.-P. Gaspard,
  T.~Irifune, J.-Y. Raty, Universal amorphous-amorphous transition in ge x se
  100- x glasses under pressure, Scientific reports 6 (2016) 27317.

\bibitem{boolchand2018topological}
P.~Boolchand, M.~Bauchy, M.~Micoulaut, C.~Yildirim, Topological phases of
  chalcogenide glasses encoded in the melt dynamics (phys. status solidi b
  6/2018), physica status solidi (b) 255~(6) (2018) 1870122.

\bibitem{gueguen2015}
Y.~Gueguen, V.~Keryvin, T.~Rouxel, M.~L. Fur, H.~Orain, B.~Bureau,
  C.~Boussard-Plédel, J.-C. Sangleboeuf,
  \href{http://www.sciencedirect.com/science/article/pii/S0167663615000605}{A
  relationship between non-exponential stress relaxation and delayed elasticity
  in the viscoelastic process in amorphous solids: Illustration on a
  chalcogenide glass}, Mechanics of Materials 85 (2015) 47 -- 56.
\newblock \href
  {http://dx.doi.org/https://doi.org/10.1016/j.mechmat.2015.02.013}
  {\path{doi:https://doi.org/10.1016/j.mechmat.2015.02.013}}.
\newline\urlprefix\url{http://www.sciencedirect.com/science/article/pii/S0167663615000605}

\bibitem{SulfurPRL}
T.~Scopigno, S.~N. Yannopoulos, F.~Scarponi, K.~S. Andrikopoulos, D.~Fioretto,
  G.~Ruocco,
  \href{https://link.aps.org/doi/10.1103/PhysRevLett.99.025701}{Origin of the
  $\ensuremath{\lambda}$ transition in liquid sulfur}, Phys. Rev. Lett. 99
  (2007) 025701.
\newblock \href {http://dx.doi.org/10.1103/PhysRevLett.99.025701}
  {\path{doi:10.1103/PhysRevLett.99.025701}}.
\newline\urlprefix\url{https://link.aps.org/doi/10.1103/PhysRevLett.99.025701}

\bibitem{Zhou2018}
T.~Zhou, Q.~Zhou, J.~Xie, X.~Liu, X.~Wang, H.~Ruan,
  \href{https://ceramics.onlinelibrary.wiley.com/doi/abs/10.1111/ijag.12290}{Elastic-viscoplasticity
  modeling of the thermo-mechanical behavior of chalcogenide glass for aspheric
  lens molding}, International Journal of Applied Glass Science 9~(2)
  252--262.
\newblock \href
  {http://arxiv.org/abs/https://ceramics.onlinelibrary.wiley.com/doi/pdf/10.1111/ijag.12290}
  {\path{arXiv:https://ceramics.onlinelibrary.wiley.com/doi/pdf/10.1111/ijag.12290}},
  \href {http://dx.doi.org/10.1111/ijag.12290} {\path{doi:10.1111/ijag.12290}}.
\newline\urlprefix\url{https://ceramics.onlinelibrary.wiley.com/doi/abs/10.1111/ijag.12290}

\bibitem{Sen}
W.~Zhu, B.~G. Aitken, S.~Sen,
  \href{https://doi.org/10.1063/1.5022787}{Communication: Observation of
  ultra-slow relaxation in supercooled selenium and related glass-forming
  liquids}, The Journal of Chemical Physics 148~(11) (2018) 111101.
\newblock \href {http://arxiv.org/abs/https://doi.org/10.1063/1.5022787}
  {\path{arXiv:https://doi.org/10.1063/1.5022787}}, \href
  {http://dx.doi.org/10.1063/1.5022787} {\path{doi:10.1063/1.5022787}}.
\newline\urlprefix\url{https://doi.org/10.1063/1.5022787}

\bibitem{boon1991molecular}
J.~P. Boon, S.~Yip, Molecular hydrodynamics, Courier Corporation, 1991.

\bibitem{trachenko2015collective}
K.~Trachenko, V.~Brazhkin, Collective modes and thermodynamics of the liquid
  state, Reports on Progress in Physics 79~(1) (2015) 016502.

\bibitem{huerta2015collective}
A.~Huerta, T.~Bryk, A.~Trokhymchuk, Collective excitations in 2d hard-disc
  fluid, Journal of colloid and interface science 449 (2015) 357--363.

\bibitem{bannerman2011dynamo}
M.~N. Bannerman, R.~Sargant, L.~Lue, Dynamo: a free $\backslash$calo (n)
  general event-driven molecular dynamics simulator, Journal of computational
  chemistry 32~(15) (2011) 3329--3338.

\bibitem{speedy1999glass}
R.~J. Speedy, Glass transition in hard disc mixtures, The Journal of chemical
  physics 110~(9) (1999) 4559--4565.

\bibitem{russo2017disappearance}
J.~Russo, N.~B. Wilding, Disappearance of the hexatic phase in a binary mixture
  of hard disks, Physical review letters 119~(11) (2017) 115702.

\bibitem{baggioli2018maxwell}
M.~Baggioli, K.~Trachenko, Maxwell interpolation and close similarities between
  liquids and holographic models, arXiv preprint arXiv:1808.05391.

\bibitem{baggioli2018solidity}
M.~Baggioli, K.~Trachenko, Solidity of liquids: How holography knows it, arXiv
  preprint arXiv:1807.10530.

\bibitem{yang2017emergence}
C.~Yang, M.~Dove, V.~Brazhkin, K.~Trachenko, Emergence and evolution of the k
  gap in spectra of liquid and supercritical states, Physical review letters
  118~(21) (2017) 215502.

\bibitem{naumis2006variation}
G.~G. Naumis, Variation of the glass transition temperature with rigidity and
  chemical composition, Physical Review B 73~(17) (2006) 172202.

\bibitem{glasstone1941theory}
S.~Glasstone, H.~Eyring, K.~J. Laidler, The theory of rate processes,
  McGraw-Hill, 1941.

\bibitem{hanggi}
P.~H{\"a}nggi, P.~Talkner, M.~Borkovec, Reaction-rate theory: fifty years after
  kramers, Reviews of modern physics 62~(2) (1990) 251.

\bibitem{toledo2017short}
J.~Q. Toledo-Mar{\'\i}n, G.~G. Naumis, Short time dynamics determine glass
  forming ability in a glass transition two-level model: A stochastic approach
  using kramers’ escape formula, The Journal of Chemical Physics 146~(9)
  (2017) 094506.

\bibitem{toledo2018escape}
J.~Q. Toledo-Mar{\'\i}n, G.~G. Naumis, Escape time, relaxation, and sticky
  states of a softened henon-heiles model: Low-frequency vibrational mode
  effects and glass relaxation, Physical Review E 97~(4) (2018) 042106.

\bibitem{gutzow1995vitreous}
I.~Gutzow, J.~Schmelzer, The Vitreous State: Thermodynamics, Structure,
  Rheology, and Crystallization, Springer-Verlag, Berlin, 1995.

\bibitem{naumistails}
G.~Naumis, G.~Cocho, The tails of rank-size distributions due to multiplicative
  processes: from power laws to stretched exponentials and beta-like functions,
  New Journal of Physics 9~(8) (2007) 286.

\bibitem{NaumisPhillips2012}
G.~Naumis, J.~Phillips,
  \href{http://www.sciencedirect.com/science/article/pii/S0378437112001215}{Diffusion
  of knowledge and globalization in the web of twentieth century science},
  Physica A: Statistical Mechanics and its Applications 391~(15) (2012) 3995 --
  4003.
\newblock \href {http://dx.doi.org/https://doi.org/10.1016/j.physa.2012.02.005}
  {\path{doi:https://doi.org/10.1016/j.physa.2012.02.005}}.
\newline\urlprefix\url{http://www.sciencedirect.com/science/article/pii/S0378437112001215}

\bibitem{sudduth2005}
R.~D. Sudduth,
  \href{https://onlinelibrary.wiley.com/doi/abs/10.1002/pen.20380}{New
  description of viscoelasticity that can be applied to mechanical properties
  such as constant strain rate, creep, and stress relaxation analysis}, Polymer
  Engineering \& Science 45~(12)  1600--1605.
\newblock \href
  {http://arxiv.org/abs/https://onlinelibrary.wiley.com/doi/pdf/10.1002/pen.20380}
  {\path{arXiv:https://onlinelibrary.wiley.com/doi/pdf/10.1002/pen.20380}},
  \href {http://dx.doi.org/10.1002/pen.20380} {\path{doi:10.1002/pen.20380}}.
\newline\urlprefix\url{https://onlinelibrary.wiley.com/doi/abs/10.1002/pen.20380}

\bibitem{MauroProny}
J.~C. Mauro, Y.~Z. Mauro,
  \href{http://www.sciencedirect.com/science/article/pii/S0378437118304795}{On
  the prony series representation of stretched exponential relaxation}, Physica
  A: Statistical Mechanics and its Applications 506 (2018) 75 -- 87.
\newblock \href {http://dx.doi.org/https://doi.org/10.1016/j.physa.2018.04.047}
  {\path{doi:https://doi.org/10.1016/j.physa.2018.04.047}}.
\newline\urlprefix\url{http://www.sciencedirect.com/science/article/pii/S0378437118304795}

\bibitem{AbyzovZanotto}
A.~S. Abyzov, V.~M. Fokin, A.~M. Rodrigues, E.~D. Zanotto, J.~W. Schmelzer,
  \href{http://www.sciencedirect.com/science/article/pii/S0022309315302325}{The
  effect of elastic stresses on the thermodynamic barrier for crystal
  nucleation}, Journal of Non-Crystalline Solids 432 (2016) 325 -- 333.
\newblock \href
  {http://dx.doi.org/https://doi.org/10.1016/j.jnoncrysol.2015.10.029}
  {\path{doi:https://doi.org/10.1016/j.jnoncrysol.2015.10.029}}.
\newline\urlprefix\url{http://www.sciencedirect.com/science/article/pii/S0022309315302325}

\bibitem{grassia2006constitutive}
L.~Grassia, A.~D’Amore, Constitutive law describing the phenomenology of
  subyield mechanically stimulated glasses, Physical Review E 74~(2) (2006)
  021504.

\bibitem{grassia2009interplay}
L.~Grassia, A.~D'Amore, On the interplay between viscoelasticity and structural
  relaxation in glassy amorphous polymers, Journal of Polymer Science Part B:
  Polymer Physics 47~(7) (2009) 724--739.

\bibitem{grassia2009relative}
L.~Grassia, A.~D’Amore, The relative placement of linear viscoelastic
  functions in amorphous glassy polymers, Journal of Rheology 53~(2) (2009)
  339--356.

\bibitem{grassia2011isobaric}
L.~Grassia, A.~D'Amore, Isobaric and isothermal glass transition of pmma:
  Pressure-volume-temperature experiments and modelling predictions, Journal of
  Non-Crystalline Solids 357~(2) (2011) 414--418.

\bibitem{grassia2011modeling2}
L.~Grassia, M.~G.~P. Carbone, G.~Mensitieri, A.~D’Amore, Modeling of density
  evolution of pla under ultra-high pressure/temperature histories, Polymer
  52~(18) (2011) 4011--4020.

\bibitem{ramos1984isobaric}
A.~Ramos, J.~Hutchinson, A.~Kovacs, Isobaric thermal behavior of glasses during
  uniform cooling and heating. iii. predictions from the multiparameter kahr
  model, Journal of Polymer Science: Polymer Physics Edition 22~(9) (1984)
  1655--1696.

\bibitem{mckenna2012physical}
G.~B. McKenna, Physical aging in glasses and composites, in: Long-term
  durability of polymeric matrix composites, Springer, 2012, pp. 237--309.

\bibitem{grassia2011modeling}
L.~Grassia, M.~G.~P. Carbone, A.~D'Amore, Modeling of the isobaric and
  isothermal glass transitions of polystyrene, Journal of Applied Polymer
  Science 122~(6) (2011) 3751--3756.

\end{thebibliography}


\bibliographystyle{elsarticle-num} 





\end{document}